\documentclass[aps,prl,twocolumn,showpacs,superscriptaddress,floatfix]{revtex4}
\usepackage{graphicx}
\begin{document}

\title{Indications of a shallow potential in 
$^{48}$Ca+$^{96}$Zr fusion reactions}
\author{H. Esbensen and C. L. Jiang}
\affiliation{Physics Division, Argonne National Laboratory, Argonne, Illinois 60439}
\date{\today}
\begin{abstract}
Fusion data for $^{48}$Ca+$^{96}$Zr are analyzed by coupled-channels 
calculations. Puzzling features of a previous analysis are eliminated by
applying a potential that has a shallow pocket in the entrance channel.
Thus the observed $S$ factor for fusion, which develops a maximum at
low energy, can be reproduced fairly well. 
The high-energy data can also be accounted for but that requires the 
use of a weak, short-ranged imaginary potential that absorbs the 
incoming flux near the location of the minimum of the potential pocket.  
Predictions of the fusion hindrance in other Ca+Zr systems are made 
and are compared with the systematics that has been developed previously.
\end{abstract}
\maketitle

\section{Introduction}

In the field of heavy-ion fusion reactions two major issues have
been discussed in recent years.
One issue is how to explain the hindrance of fusion at low energies, 
that is the steep falloff of the measured fusion cross sections, 
which has been observed in many heavy-ion systems at energies far 
below the Coulomb barrier 
(see Ref. \cite{jiangsys} and references therein.)  
The hindrance is often so strong that the $S$ factor for fusion 
develops a maximum at low energy, and it has been a challenge to 
reproduce this behavior in coupled-channels calculations.
The other issue is the suppression of high-energy fusion data, 
which is observed when the data are compared to conventional
one-dimensional or coupled-channels calculations \cite{newton}.

A practical solution to the problems mentioned above has been 
to adjust the parameters of the Woods-Saxon shaped, ion-ion 
potential that is used in the calculations.
Thus it has been realized that by using a large diffuseness 
(in combination with a deep potential and a small radius parameter)
one can often resolve the problem of a suppression at high energy 
\cite{newton}.
However, the diffuseness that is required to reproduce the high-energy 
data, for example, in the fusion of $^{16}$O+$^{208}$Pb , is different 
from the value that is needed to fit the low-energy data \cite{nanda}.
Moreover, the values are often unrealistic and much larger than the 
empirical diffuseness which has been extracted from analyses of 
elastic scattering data or obtained from double-folding 
potentials \cite{BW}.  It has recently been shown that a large 
diffuseness is also inconsistent with quasi-elastic scattering data 
\cite{evers}. 

A good example on a system where both of the above mentioned problems 
exist is the fusion data for $^{48}$Ca+$^{96}$Zr \cite{stefca48}.
The problems were solved in the original analysis of the data by using 
a large diffuseness, $a$ = 0.85 fm, instead of the empirical value, 
$a$ = 0.677 fm (Eq. (III.44) of Ref. \cite{BW}). 
The measurements \cite{stefca48} were later supplemented with two new 
low-energy data points \cite{trotta} and they confirmed that the $S$ 
factor for this system does develop a maximum at low energy.

Instead of using an ion-ion potential with an anomalously large diffuseness, 
the double-folding technique is applied to construct an ion-ion potential 
which has a realistic behavior at large distances and is adjusted at small 
distances between the reacting nuclei to produce a shallow pocket in the 
entrance channel potential \cite{misc75}.
The potential used is the so-called M3Y+repulsion potential and the 
construction of it is described in Ref.  \cite{misc75}. 
The basic feature of this potential is that the existence of a shallow
pocket in the entrance channel makes it possible to explain the steep 
falloff of the fusion data at low energies, and it also helps explaining the 
measured fusion cross sections at high energies \cite{misc75,misopb,esboo}. 
However, in order to reproduce the high-energy data, it is often 
necessary to include a short-ranged, imaginary potential in the calculations 
of the cross sections \cite{misopb,esboo}. 

The M3Y+repulsion potential is applied in coupled-channels calculations
of the fusion cross section for $^{48}$Ca+$^{96}$Zr. The purpose is to 
demonstrate that the energy dependence of the measured $S$ factor for 
fusion can be reproduced at low energies by employing this potential.  
It is also shown that the calculated cross sections at high energies are 
sensitive to the nuclear potential and to the nuclear structure input,
and it is possible to reproduce the high-energy data by using a realistic 
structure input and employing the M3Y+repulsion potential, 
in combination with a short-ranged imaginary potential.

Finally, the hindrance in the fusion of other Ca+Zr isotopes at low energies 
is discussed and compared with the systematics we have developed
previously \cite{jiangsys}, and predictions are made of the energy 
where the $S$ factor for fusion is expected to develop a maximum 
in those cases where it has not yet been observed experimentally.

\section{Ion-ion potential}

The construction of the double-folding ion-ion potential for 
$^{48}$Ca+$^{96}$Zr is based on the effective M3Y interaction which 
is supplemented with a contact term for the exchange part \cite{misc75}.
The densities are parametrized as fermi functions with the radius 
3.675 fm and diffuseness 0.525 fm for $^{48}$Ca, and the radius 
5.02 fm and diffuseness 0.55 fm for $^{96}$Zr.
These two densities produce the RMS matter radii of 3.45 and 4.39 fm,
respectively, which are consistent with the measured charge radii.

\begin{figure}
\includegraphics[width = 8cm]{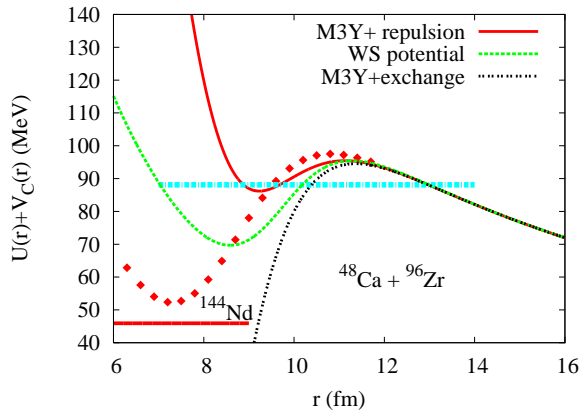}
\caption{\label{pot4896} (color online)
The M3Y+repulsion potential (red solid curve) 
is compared to the Woods-Saxon potential (middle dashed curve) 
and to the pure M3Y+exchange potential (lowest black dashed curve). 
The diamonds show the potential with large diffuseness that 
was used in Ref. \cite{stefca48}.
The energy of the compound nucleus $^{144}$Nd and the energy band 
where the $S$ factor for fusion has a maximum ($\approx$ 88 MeV) 
are also indicated.}
\end{figure}

The M3Y double-folding potential (including exchange) is illustrated 
by the lowest dashed curve in Fig. \ref{pot4896}. It is unrealistic at 
short distances between the two nuclei because it is much deeper than
the ground state energy of the  compound nucleus. 
The M3Y interaction has therefore been corrected by supplementing it
with a repulsive term as described in Ref. \cite{misc75}.
The densities that are used in calculating the associated repulsive, 
double-fording potential have a sharper profile, with the diffuseness 
$a_{\rm rep}$=0.4 fm, but the radii are the same as used above to
calculate the ordinary M3Y double-folding potential.
The strength of the repulsive contact term, $V_{\rm rep}$ = 507 MeV,
was adjusted to produce the nuclear incompressibility $K$ = 223 MeV.

The entrance channel potential for the M3Y+repulsion
potential is illustrated by the upper solid curve in Fig. \ref{pot4896}. 
It has a rather shallow pocket inside the Coulomb barrier. 
Also shown is the entrance channel potential obtained from a Woods-Saxon 
(WS) potential, which is similar to the empirical Aky\"uz-Winther potential 
\cite{BW}.
The slightly adjusted parameters of this WS potential are
$V_0$ = -77.18 MeV, $R$ = 9.686 fm, and $a$ = 0.677 fm.
They produce the same Coulomb barrier height, $V_{CB}$ = 95.4 MeV, 
as the M3Y+repulsion potential but the pocket is deeper.
The fusion cross sections obtained in the coupled-channels calculations 
that are based on these two potentials are essentially the same at 
energies close to the Coulomb barrier but differ at energies far 
below the Coulomb barrier. This will be discussed in the next section.

The entrance channel potential which was used in Ref. \cite{stefca48} 
(with diffuseness $a$ = 0.85 fm) is indicated by the 
diamonds in Fig. \ref{pot4896}. This potential is deeper than the other 
two potentials mentioned above but none of them are deeper than the ground 
state energy of the compound nucleus $^{144}$Nd, which is indicated by the
lowest horizontal line. 
The broad horizontal band near 88 MeV indicates the energy where the
experimental $S$ factor for fusion develops a maximum.

\section{Coupled-channels calculations}

\begin{table}
\caption{Structure input for $^{48}$Ca \cite{flem} and $^{96}$Zr \cite{NDS}.} 
\begin{tabular} {|c|c|c|c|c|c|}
\colrule
Nucleus & $\lambda^\pi$ &  E$_x$ (MeV) & 
$\beta_\lambda^C$ & $\sigma_\lambda^C$ (fm) 
& $\sigma_\lambda^N$ (fm) \\ 
\colrule
$^{48}$Ca   & $2^+$       & 3.832 & 0.102 & 0.126 & 0.190 \\
            & 2PH($2^+$)  & 4.297 & 0.082 & 0.101 & 0.101 \\
            & $3^-$       & 4.507 & 0.203 & 0.250 & 0.190 \\
\colrule
 $^{96}$Zr  & $2^+$       & 1.751 & 0.079 & 0.123 & 0.123 \\ 
            & $3^-$       & 1.897 & 0.295 & 0.457 & 0.457 \\ 
\colrule
\end{tabular}
\end{table}

The nuclear structure input to the coupled-channels calculations  
associated with the excitation of the $2^+$ and $3^-$ states in 
projectile and target is shown in Table I.  
The calculations include all one-phonon, 
two-phonon, and mutual excitations of these one-phonon excitations. 
That is a total of 15 channels in the rotating frame approximation
\cite{misc75}. The coupling between the one- and two-phonon 
states is assumed to be vibrational (harmonic), except for the 
quadrupole excitation of $^{48}$Ca for which we use a weaker one-
to two-phonon coupling strength (see Table I) which is based on 
the known $0_2^+\rightarrow 2_1^+$ transition strength \cite{NDS}.

The conventional assumption $\beta^N=\beta^C$ does not work so well 
for $^{48}$Ca. The empirical nuclear coupling strengths that will 
be used are shown in the last column of Table I. They are expressed 
in terms of the values of 
\begin{equation}
\sigma_\lambda^N=\frac{\beta_\lambda^N R_N}{\sqrt{4\pi}},
\label{zpm}
\end{equation}
and were determined by analyzing measurements of the inelastic 
scattering of $^{16}$O on $^{48}$Ca \cite{flem}.

\begin{figure}
\includegraphics[angle=270,width = 8cm]{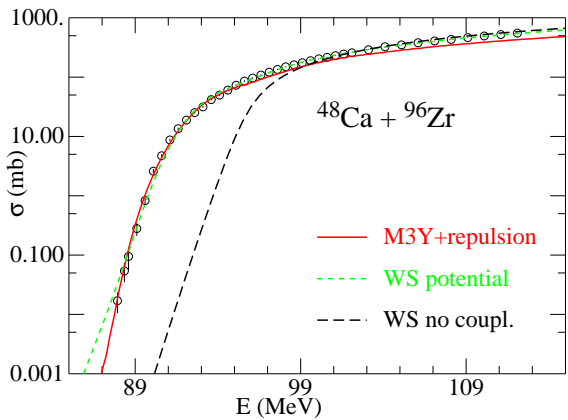}
\caption{ (color online)
Measured fusion cross sections for$^{48}$Ca+$^{96}$Zr \cite{stefca48,trotta} 
are compared to coupled-channels calculations
based on the Woods-Saxon (WS, green dashed curve) and M3Y+repulsion 
(red solid curve) potentials.
The lowest dashed (black) curve is the no-coupling limit for the 
WS potential.}
\label{4896f}
\end{figure}

The fusion cross sections are determined by imposing ingoing-wave 
boundary conditions at the position of the minimum of pocket in the 
entrance potential. The results of coupled-channels calculations are 
compared in Fig. \ref{4896f} to the data of Refs. \cite{stefca48,trotta}. 
The top dashed (green) curve is based on the empirical Woods-Saxon 
potential discussed in the previous section. It is seen to reproduce
the data quite well, except at the very lowest energy.
The overall $\chi^2/N$ (based on statistical errors only) is about 
4 for this calculation.

The solid (red) curve in Fig. \ref{4896f} is the coupled-channels 
result obtained with the M3Y+repulsion potential. 
It reproduces the data very well at low energies but is below
the data at high energies. The latter problem can be solved as
mentioned earlier by employing short-ranged imaginary potential.
The explicit form used below is
\begin{equation}
W(r) = \frac{-10 \ {\rm MeV}}{1+\exp((r-r_{p})/a_w)}.
\label{w(r)}
\end{equation}
where $r_p$ is the radial distance between the reacting nuclei at the 
minimum of the pocket, and the diffuseness $a_w$ is set to 0.2 fm. 
The effect of the imaginary potential, Eq. (\ref{w(r)}), is illustrated 
below (see Fig. \ref{4896lf}.)


\subsection{Barrier distribution}

The effect of couplings on the calculated fusion cross section is 
very strong. This can be seen by comparing the coupled-channels 
calculations to the no-coupling limit which is shown by the lower 
dashed curve in Fig. \ref{4896f}.  The effect is to shift the 
calculated cross section by roughly 4--5 MeV to lower energies. 

\begin{figure}
\includegraphics[width = 8cm]{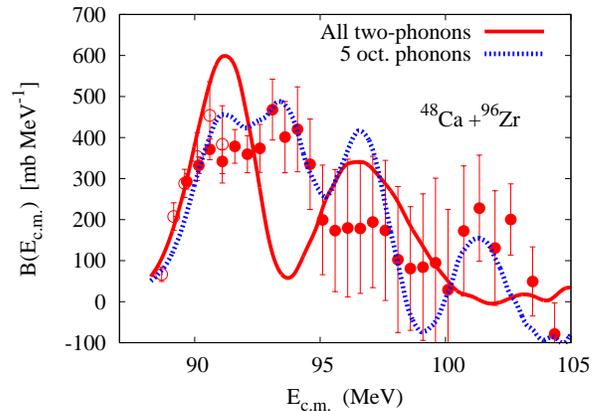}
\caption{\label{48962d} 
(color online)
Barrier distributions obtained from the fusion 
cross sections shown in Fig. \ref{4896f} using $\Delta E$ = 2 MeV
(open circles for $\Delta E$ = 1 MeV).
The red solid curve was derived from the M3Y+repulsion calculation
show in Fig. \ref{4896f}. The blue, dashed curve was obtained from
a calculation that includes up to 5 phonons of the octupole 
excitation in $^{96}$Zr.}
\end{figure}

The couplings produce some interesting features in the barrier 
distribution for fusion \cite {rowley}, which is defined as the 
second derivative of the energy-weighted fusion cross section 
with respect to the center-of-mass energy $E$,
\begin{equation}
\label{barrier}
B((E) = \frac{d^2(E\sigma_f)}{dE^2}.
\end{equation}
The barrier distributions, which are shown in Fig. \ref{48962d}, were 
calculated by the finite difference method using the same energy step 
$\Delta E$ = 2 MeV that was used in the earlier work \cite{stefca48}.
The measured distribution is broad and contains several peak structures. 
The red solid curve in Fig. \ref{48962d} is the distribution derived 
from the coupled-channels calculations that are based on the 
M3Y+repulsion potential. 
It is broad and consists of two strong peaks but the shape is not in 
good agreement with the data. 

It was demonstrated in Ref.  \cite{stefca48} that some of these peak 
structures in the measured barrier distribution can be reproduced by 
including up to three-phonon excitations of the octupole mode in $^{96}$Zr.
The blue dashed curve in Fig. \ref{48962d} is the barrier distribution 
one obtains by considering up to five phonon excitations of the octupole 
mode in $^{96}$Zr in the coupled-channels calculations.
The other excitations shown in Table I were not included in these
calculations for practical reasons.
The shape of the distribution (the blue dashed curve) seems to trace 
the four peak structures of the measured distribution rather well,
and it appears to be in better agreement with the data than the 
three-phonon calculation presented in Ref.  \cite{stefca48}.
It is obvious from this comparison that the calculated barrier 
distribution is very sensitive to multiphonon excitations of 
the soft octupole mode in $^{96}$Zr. 

The fusion cross section obtained in the calculation that is based on 
up to five-phonon octupole excitations has an average deviation of 
about 10\% from the data (i.~e., we need a systematic error of 10\% 
in order to obtain a $\chi^2$ per point of 1), whereas the solid 
curve in Fig. 1 deviates on average by 7\%. 
It is possible that one would be able to improve the overall fit to the 
data by including more channels in the calculation. However, such a 
calculation would probably not be very realistic because of the sensitivity 
to the empirically poorly known multiphonon excitations. The main
point we want to make is that the standard calculation, which is based
on couplings to one- and two-phonon excitations, provides a reasonable 
fit to the data and by applying the M3Y+repulsion potential it is 
possible to reproduce the fusion hindrance observed at energies far
below the Coulomb barrier. 

\subsection{Behavior at low energy}  

\begin{figure}
\includegraphics[width = 8cm]{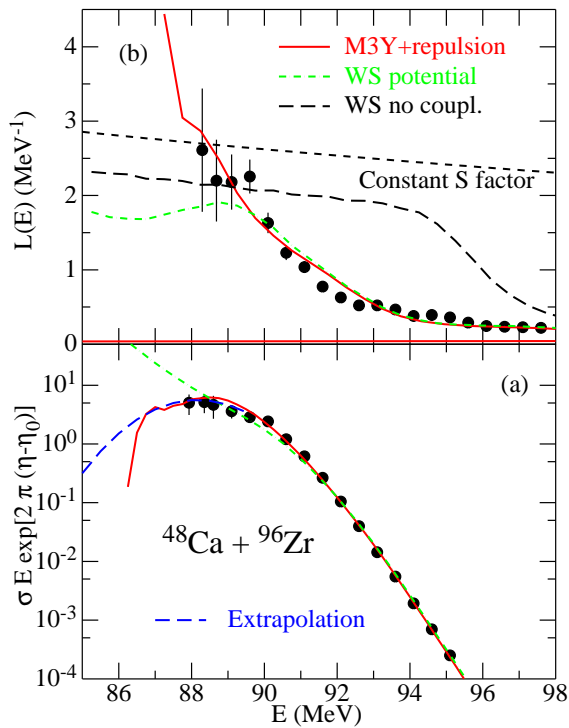}
\caption{ (color online) (a) The measured $S$ factors for the 
fusion of $^{48}$Ca+$^{96}$Zr \cite{stefca48,trotta} are compared 
to the two coupled-channels calculations shown in Fig. \ref{4896f}. 
The value of $\eta_0$ was set to $\eta_0$ = 74.58. 
The blue dashed curve is the prediction of Eq. (\ref{sige}) 
and the parameters of Table II.
The logarithmic derivatives, Eq. (\ref{log}), are shown in (b). 
The top short-dashed curve is the constant $S$ factor limit, Eq. (\ref{lcs});
the long-dashed curve is the no-coupling limit for the WS potential.}
\label{4896sf}
\end{figure}

The low-energy behavior of the fusion cross sections is illustrated
in Fig. \ref{4896sf}a in terms of the $S$ factor for fusion, which is 
defined by,
\begin{equation}
S(E) = \sigma_f(E) \ E \ \exp\Bigl(2\pi\eta(E)\Bigr),
\end{equation}
where $E$ is the center-of-mass energy, 
$\eta=Z_1Z_2e^2/(\hbar v)$ is the Sommerfeld parameter, 
and $v$ is the relative velocity. 
It is seen that the $S$ factor for the coupled-channels calculations 
that are based on the Woods-Saxon potential keeps increasing 
with decreasing energy. The calculation based on the M3Y+repulsion 
potential, on the other hand, develops a maximum before it goes to 
zero near 86.2 MeV, which is the energy of the pocket minimum for 
the M3Y+repulsion potential shown in Fig. \ref{pot4896}.
The calculated fusion cross section must go to zero at this energy 
because we use a real potential at low energy and use ingoing-wave 
boundary conditions to determine the fusion probability.

The coupled-channels calculations that are based on the M3Y+repulsion 
potential provide the best description of the low-energy $S$ factor 
data shown in Fig. \ref{4896sf}a.
In fact, the $S$ factor data have just reached a maximum at the lowest 
energies, which is a characteristic feature for the onset of the fusion 
hindrance phenomenon in systems with a negative Q-value for fusion, and 
this feature is reproduced by using a shallow entrance channel potential.
The extrapolation to very low energies, which will be discussed later 
on, is also shown in the figure. It has a nearly parabolic shape and 
the lowest data points are seen to fall on the top of this curve.

Another way to illustrate the fusion hindrance is to show the 
logarithmic derivative of the energy weighted fusion cross section,
\begin{equation}
L(E) = \frac{d\ln(E\sigma_f)}{dE}
= \frac{1}{E\sigma_f} \ \frac{d(E\sigma_f)}{dE}. 
\label{log}
\end{equation}
This quantity is shown in Fig. \ref{4896sf}b.
The calculations based on the Wood-Saxon potentials are seen to reach 
a local maximum at low energies. The data and the calculations based on 
the M3Y+repulsion potential, on the other hand, keep increasing steeply 
with decreasing energy. They both intersect the upper `constant $S$ factor' 
line, which is the logarithmic derivative one obtains by assuming a 
constant $S$ factor \cite{jiangsys},
\begin{equation}
L_{cs}(E)= \frac{\pi\eta}{E}=
\frac{0.495 Z_1Z_2\sqrt{\mu}}{E^{3/2}}.
\label{lcs}
\end{equation} 
Here $\mu=A_1A_2/(A_1+A_2)$ is the reduced mass number.
At the point where $L(E)$ intersects with $L_{cs}(E)$, the associated
$S$ factor will have a maximum (see Fig.  \ref{4896sf}a.) 

\subsection{Behavior at high energy}  

\begin{figure}
\includegraphics[angle=270,width = 8cm]{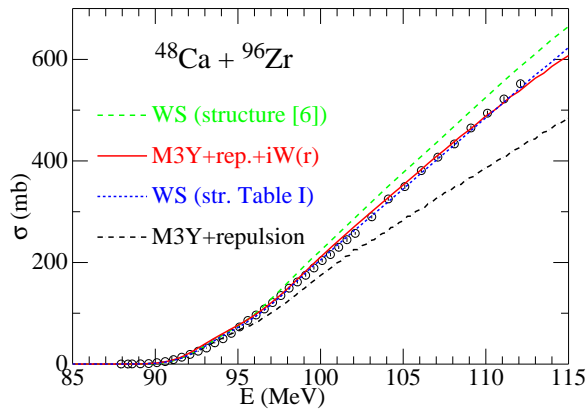}
\caption{ (color online)
Linear plot of the fusion cross sections for $^{48}$Ca+$^{96}$Zr 
\cite{stefca48}.  The upper dashed curve is based on the WS potential 
with diffuseness $a$ = 0.677 fm and the structure input from Ref. 
\cite{stefca48}. The next (blue) dashed curve is based on the same
potential but stucture input is from Table I.
The red solid and the lower dot-dashed curves are the coupled-channels
calculations, based on the M3Y+repulsion potential and the structure 
input of Table I, with and without the effect of the short-ranged 
imaginary potential $iW(r)$, respectively.} 
\label{4896lf}
\end{figure}

The behavior of the fusion cross sections at high energies is 
illustrated in Fig. \ref{4896lf}.
The lowest dashed curve is the result of the coupled-channels
calculation that is based on the M3Y+repulsion potential.
This calculation is suppressed compared to the data, as already
pointed out in the discussion of Fig. \ref{4896f}. However,
the discrepancy with the data can essentially be eliminated by 
including in the calculation the weak, short-ranged imaginary 
potential, Eq. \ref{w(r)}. This is illustrated by the red solid
curve.  
The result obtained with the Woods-Saxon potential (with diffuseness 
$a$ = 0.677 fm) and the structure input from Table I is shown by the
blue dashed curve which is difficult to see because it is essentially 
identical to the solid curve in Fig.  \ref{4896lf}.
This remarkable feature is the reason why the two calculations give
essentially the same quality fit to the data as mentioned earlier.

Good agreement with the high energy data could only be achieved in
Ref. \cite{stefca48} by using a large diffuseness in the Woods-Saxon 
potential.  This seems to contradict the results discussed above,
where the Woods-Saxon potential with `normal' diffuseness provided
an excellent description of the high-energy data when applied in 
coupled-channels calculations. 
The reason for this apparent contradiction is that the structure 
input of Table I is different from the structure input used in 
Ref. \cite{stefca48}, in particular for $^{48}$Ca.
The coupled-channels calculations were therefore repeated using the 
structure input from Ref.  \cite{stefca48} and a Woods-Saxon potential 
with the `normal' diffuseness $a$ = 0.677 fm.
The result is shown by the upper dashed curve in Fig. \ref{4896lf}. 
It is seen that the data are suppressed compared to this calculation, 
in accord with the findings in Ref. \cite{stefca48}.

\section{Fusion hindrance in Ca+Zr systems}

Many fusion reactions involving medium heavy nuclei have been 
measured at Legnaro. Here we focus on the measurements that have been
performed with different isotopes of calcium and zirconium 
\cite{stefca40,stefca48,stefzr94}. The purpose is to extract information 
about the fusion hindrance phenomenon at extreme subbarrier energies
and to compare with the results that have been obtained for other systems 
\cite{jiangsys,ninia,ninib}.

The fusion hindrance for systems with negative Q-value 
can be characterized by the energy $E_s$ where the 
$S$ factor for fusion has a maximum at low energy \cite{jiangsys}.
Although many measurements do not reach such low energies, it is still
possible to determine $E_s$ by extrapolation of the existing data.
It was found that the logarithmic derivative of the energy-weighted
fusion cross section, the $L(E)$ defined in Eq. (\ref{log}), has a nearly 
linear dependence on energy at energies around the $S$ factor maximum
\cite{jiangsys,ninib}. A slightly different (non-linear) parametrization
is used in the following \cite{impress}, namely,
\begin{equation}
L(E) = A_0 + \frac{B_0}{(E+Q)^{N_p}},
\label{leg}
\end{equation}
where $Q$ is the $Q$ value for the fusion reaction, and we assume that
$N_p$=3/2. This expressions implies that $L(E)\rightarrow\infty$ for
$E\rightarrow -Q$ in agreement with the constraint that the cross section
must go to zero when the center-of-mass energy $E$ approaches the ground
state energy, $-Q$, of the compound nucleus \cite{jiangsys,ninib}.

\begin{figure}
\includegraphics[width = 8cm]{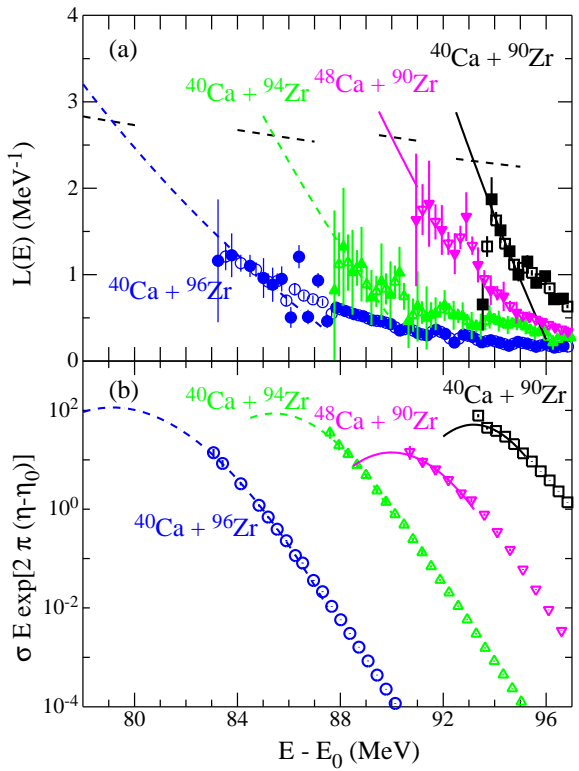}
\caption{\label{clj1} (color online)
Logarithmic derivative (a) and $S$ factors (b) for the fusion 
of $^{40}$Ca+$^{90,94,96}$Zr
\cite{stefca40,stefzr94} and $^{48}$Ca+$^{90}$Zr \cite{stefca48}.
The curves in (a) are fits to the measured $L(E)$ using the parameterization 
Eq. (\ref{leg}).  
The upper, nearly horizontal dashed lines show the constant 
$S$ factor limit, Eq. (\ref{lcs}).  
The curves in (b) are the $S$ factors predicted by
Eq. (\ref{sige}) and the parameters in Table II.
The parameters $E_0$ and $\eta_0$ used in the plot
are $E_0$ = 0, 1, 3, 0 MeV and $\eta_0$ = 68.59, 70.87, 71.71, 73.98 
for the systems $^{40}$Ca+$^{90,94,96}$Zr and $^{48}$Ca+$^{90}$Zr, 
respectively.} 
\end{figure}

Once the parameters $A_0$ and $B_0$ in Eq. \ref{leg} have been determined 
from the low energy data, it is straightforward to extrapolate and find 
the energy $E_s$ where $L(E)$ intersects the logarithmic derivative 
$L_{cs}(E)$ for a constant $S$ factor, Eq. (\ref{lcs}).
Results for the systems $^{40}$Ca+$^{90,94,96}$Zr and $^{48}$Ca+$^{90}$Zr 
using this extrapolation method are shown in Fig. \ref{clj1}a. 
The parameters obtained by fitting the low-energy $L(E)$ data and the
extrapolated values of $E_s$ are collected in Table II. There we also give 
the value $L_s$ which is the logarithmic derivative at the energy $E_s$.

\begin{table} 
\caption{
Parameters for the fusion hindrance behavior, 
$E_s, L_s, A_0, B_0 $ and $\sigma_s$, 
obtained for the Ca+Zr systems.
The last column is the reference energy $E_s^{\rm ref}$ 
defined in Eq. (\ref{eref}). 
}
\begin{tabular}{|c|cccccc|}
\hline
 System                & $E_s$ & $L_s$ & $A_0$ & $B_0$
  & $\sigma_s$ & $E_s^{\rm ref}$ \\ 
          &   MeV & MeV$^{-1}$ & MeV$^{-1}$ & MeV$^{1/2}$ & mb & MeV \\ 
   \hline
   \hline
 $^{40}$Ca+$^{90}$Zr& 93.2 & 2.31 & -17.1 & 4222 & 0.37   & 92.8 \\ 
 $^{40}$Ca+$^{94}$Zr& 86.4 & 2.61 & -11.6 & 3585 & 3.8E-3 & 93.2 \\ 
 $^{40}$Ca+$^{96}$Zr& 83.1 & 2.77 & -7.43 & 2783 & 3.7E-5 & 93.4 \\ 
 $^{48}$Ca+$^{90}$Zr& 90.0 & 2.59 & -12.9 & 3775 & 0.025  & 96.7 \\ 
 $^{48}$Ca+$^{96}$Zr& 88.1 & 2.71 & -13.8 & 4528 & 0.030  & 97.4 \\ 
   \hline
\end{tabular}
\label{table2}
\end{table}

The analytic expression for the cross section one obtains from
the parametrization (\ref{leg}) of $L(E)$ is,
$$\sigma(E) = \sigma_s \ \frac{E_s}{E} \ \exp\Bigl(A_0(E-E_s)$$ 
\begin{equation}
- \frac{B_0}{(N_p-1)(E_s+Q)^{N_p-1}} \
\Bigl[\bigl(\frac{E_s+Q}{E+Q}\bigr)^{N_p-1}-1\Bigr]\Bigr).
\label{sige}
\end{equation}
The only unknown parameter in this expression is the normalization factor 
$\sigma_s$ which is the fusion cross section at the energy $E_s$.
It can be obtained by normalizing to the low-energy fusion data and
the resulting values are given in Table II. 
The $S$ factors that have been determined in this way are shown by the 
curves in Fig.  \ref{clj1}b. 
They have a characteristic, nearly parabolic behavior at 
low energy. The measured $S$ factors have not  reached a maximum for 
most of these systems but the data for $^{40}$Ca+$^{90}$Zr are close 
(although the lowest point has a somewhat unusual behavior.)
The $S$ factor for the system $^{48}$Ca+$^{96}$Zr, which was already
shown in Fig. \ref{4896sf}a, did reach a maximum at the lowest energy,
and the extrapolation obtained from the parameters in Table II
was also shown there.

\begin{figure}
\includegraphics[angle=270,width = 8cm]{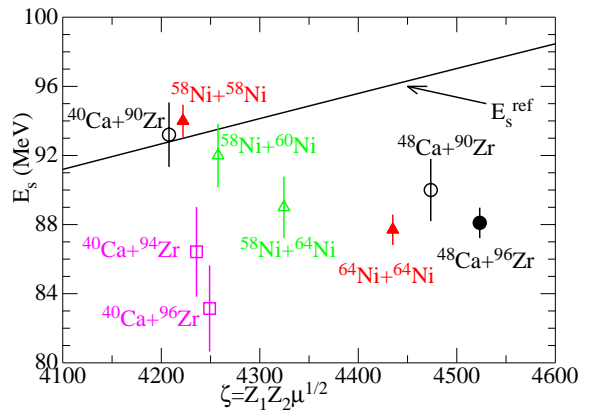}
\caption{\label{clj2} The energy $E_s$ 
for Ca+Zr and Ni+Ni systems. The solid curve is the reference
value, Eq. (\ref{eref}).
The open symbols are extrapolated values. The solid symbols are for systems 
where a maximum in the $S$ factor has been observed.}
\end{figure}

We have previously studied the systematics of the energy $E_s$ 
for other heavy-ion systems (Refs. \cite{jiangsys,ninib}) and found 
that the value of $L_s$, i.~e., the logarithmic derivative of the 
energy weighted cross section at the energy $E_s$, has a nearly constant 
value of $L_s^{\rm ref}$ = 2.33 MeV$^{-1}$ for systems that consist of 
closed shell nuclei or of nuclei that are relatively stiff. 
Inserting such a constant value of $L_s$ and the energy $E_s$ into 
Eq. (\ref{lcs}), it is seen that the system dependence of $E_s$ is 
governed by the entrance channel parameter
$$
\zeta=Z_1Z_2\sqrt{\mu} = Z_1Z_2\sqrt{A_1A_2/(A_1+A_2)}.
$$
The expression one obtains for $E_s$ is the reference energy
\begin{equation}
E_s^{\rm ref} = (0.495 \zeta /L_s^{\rm ref})^{2/3} \ ({\rm MeV}),
\label{eref}
\end{equation}
which is illustrated by the solid curve in Fig. \ref{clj2}.
For systems consisting of soft or open shell nuclei, the value
of $E_s$ is usually lower than the reference value $E_s^{\rm ref}$,
The deviation of $E_s$ from $E_s^{\rm ref}$ is a measure of 
the softness the systems \cite{ninia,ninib,nimo}.

The values of $E_s$ extracted from the data for the different 
Ca+Zr systems \cite{stefca48,trotta,stefca40,stefzr94} are shown in Fig. 
\ref{clj2}. Also shown are the values obtained previously for the
Ni+Ni systems, consisting of different nickel isotopes.
It is seen that the extrapolated value of $E_s$ for $^{40}$Ca+$^{90}$Zr 
falls on the solid line, consistent with the closed shell nature of these 
two nuclei. The results for the two soft and open shell systems, 
$^{48}$Ca+$^{96}$Zr and $^{64}$Ni+$^{64}$Ni, are similar and fall far 
below the solid curve.
The value of $E_s$ for $^{48}$Ca+$^{96}$Zr, for example, is 9.3 MeV 
below the reference value $E_s^{\rm ref}$, according to Table II. 

The values of $E_s$ are also far below the reference value $E_s^{\rm ref}$
for some of the other Ca+Zr systems. This is partly caused by the softness 
of the reacting nuclei or by transfer reactions.
Stefanini et al. in Ref.  \cite{stefzr94} pointed out that neutron transfer 
with positive Q value should affect the fusion of $^{40}$Ca+$^{94}$Zr and 
$^{40}$Ca+$^{96}$Zr and enhance it below the Coulomb barrier.  
Consequently, the fusion hindrance would be pushed to lower energies.

There is a correlation between the ratio $E_s/E_s^{\rm ref}$ and the 
number of valence neutrons in the two reacting nuclei, which was first
pointed out in Ref.  \cite{nimo} for nickel induced fusion reactions.
A similar trend is observed in Fig. \ref{clj2} for the Ca+Zr systems.
In order to see the interplay between neutron transfer with positive 
Q value and the fusion hindrance behavior, it would be very interesting 
to measure the fusion of $^{40}$Ca+$^{94}$Zr and $^{40}$Ca+$^{96}$Zr
down to much smaller cross sections. 

\section{Conclusions}

We conclude that there are strong indications of a fusion 
hindrance at low energy in the data for $^{48}$Ca+$^{96}$Zr. 
The $S$ factor does reach a maximum at the lowest energies and 
there is a preference among the coupled-channels calculations 
presented here for those that are based on a shallow potential 
in the entrance channel. 
In this respect, the fusion data exhibit the same characteristics
that have been observed in many other heavy-ion systems 
\cite{jiangsys,misc75,misopb,esboo}.

The systematics of the fusion hindrance in other Ca+Zr systems, 
where it has not yet been confirmed experimentally, was discussed
and predictions were made of the energy $E_s$ where the $S$ factor for 
fusion is expected to develop a maximum.
This energy can be far below the energy predicted by the empirical
formula, which was developed for systems consisting of stiff or closed 
shell nuclei. The deviation of $E_s$ from the empirical formula seems 
to correlate with the number of valence neutrons in the reacting nuclei,
which may reflect an increased softness or a stronger effect of neutron 
transfer.

The behavior of the high-energy fusion data can also be reproduced 
by applying the shallow potential in the coupled-channels calculations 
but that requires an additional weak absorption that acts near the 
minimum of the potential pocket.
The calculated, high-energy cross sections are sensitive to the nuclear 
structure input parameters. By applying the structure input which 
has been extracted from inelastic scattering data it is possible to 
reproduce the high-energy data, even with a Woods-Saxon potential that 
has a normal diffuseness. 

The energy dependence of the fusion cross section in the barrier region
was also investigated.  This region is best studied in terms of the 
extracted barrier distribution, which is usually very sensitive to the 
low-lying nuclear structure input in the coupled-channels calculations.
The calculated barrier distribution for $^{48}$Ca+$^{96}$Zr turned out
to be very sensitive to multi-phonon excitations, in particular of 
the soft octupole node in $^{96}$Zr.
The best agreement with the measured barrier distribution was achieved 
by including excitations up to five phonons of this excitation mode. 
Unfortunately, the empirical knowledge of the nuclear structure at such 
high excitation energies is poor and many other reaction channels may 
also affect the barrier distribution. 
It is therefore a serious challenge to theory to determine a credible 
nuclear structure input and explain the measured barrier distribution 
in detail.


{\bf Acknowledgments}. 
We are grateful to A. Stefanini for many communications 
and for providing us with the data, and to B. B. Back, R. V. F. Janssens,
and K. E. Rehm for a long-term collaboration on the subject of fusion 
hindrance. This work was supported by the U.S. Department of Energy, 
Office of Nuclear Physics, under Contract No. DE-AC02-06CH11357.


\begin{thebibliography}{99}
\bibitem{jiangsys} C. L. Jiang, B. B. Back, H. Esbensen, R. V. F. Janssens, and
K. E. Rehm, Phys. Rev. {\bf 73}, 014613 (2006).
\bibitem{newton} J. O. Newton {\it et al.}, Phys. Lett. {\bf B586}, 219 (2004).
\bibitem{nanda} M. Dasgupta {\it et al.}, Phys. Rev. Lett. {\bf 99}, 192701 (2007).
\bibitem{BW} R. A. Broglia and Aa. Winther, {\it Heavy-ion Reactions},
Lecture Notes, Addison-Wesley (Redwood City CA, 1991).
\bibitem{evers} M. Evers {\it et al.}, Phys. Rev. C {\bf 78}, 034614 (2008).
\bibitem{stefca48} A. M. Stefanini {\it et al}., Phys. Rev. C {\bf 73}, 034606 (2006).
\bibitem{trotta} M. Trotta {\it et al.}, Nucl. Phys. {\bf A787}, 134c ( 2007).
\bibitem{misc75} S. Mi\c sicu and H. Esbensen, Phys. Rev. C {\bf 75}, 034606 (2007).
\bibitem{misopb} H. Esbensen and \c S. Mi\c sicu, Phys. Rev. C {\bf 76}, 054609 (2007).
\bibitem{esboo} H. Esbensen, Phys. Rev. C {\bf 77}, 054608 (2008).
\bibitem{flem} H. Esbensen and F. Videb{\rm\char'32}k, Phys. Rev. C {\bf 40}, 126 (1989).
\bibitem{NDS} Evaluated Nuclear Data Structure Files (ENSDF),
National Nucl. Data Center, Brookhaven Nat. Lab., http://www.nndc.bn.gov/.
\bibitem{rowley} N. Rowley, G. R. Satchler, and P. H. Stelson, 
Phys. Lett. B {\bf 254}, 25 (1991).
\bibitem{multi} H.  Esbensen, Phys. Rev. C {\bf 72}, 054607 (2005).
\bibitem{stefca40} H. Timmers {\it et al}., Nucl. Phys. {\bf A633}, 421 (1998).
\bibitem{stefzr94} A. M. Stefanini {\it et al}., Phys. Rev. C {\bf 76}, 014610 (2007).
\bibitem{ninia} C. L. Jiang {\it et al.}, Phys. Rev. Lett. {\bf 93}, 012701 (2004).
\bibitem{ninib} C. L. Jiang, H. Esbensen, B. B. Back, R. V. F. Janssens, K. E. Rehm, Phys. Rev. C {\bf 69}, 014604 (2004).
\bibitem{impress} C. L. Jiang, K. E. Rehm, B. B. Back, and R. V. F. Janssens,
Phys. Rev. C (in press).
\bibitem{nimo} C. L. Jiang {\it et al.}, Phys. Rev. C {\bf 71}, 044613 (2005).
\end{thebibliography}
\end{document}